\documentstyle[aps,preprint]{revtex}
\begin{document}
\draft
\preprint{SOGANG-HEP-224/97}
\title{Duality of Quasilocal Black Hole 
Thermodynamics} 
\author{
Jeongwon Ho\footnote{E-mail address : 
jwho@physics3.sogang.ac.kr},
Won Tae Kim\footnote{E-mail address : 
wtkim@ccs.sogang.ac.kr}, and 
Young-Jai Park\footnote{E-mail address : 
yjpark@ccs.sogang.ac.kr}
}
\address{
Department of Physics and Basic Science 
Research Institute,\\
Sogang University, C.P.O.Box 1142, Seoul 
100-611, Korea\\
}
\date{November 1997}
\maketitle
\begin{abstract}
We consider T-duality of the quasilocal black hole
thermodynamics for the three-dimensional low energy effective
string theory. Quasilocal thermodynamic variables in the 
first law are explicitly calculated on a general
axisymmetric three-dimensional black hole solution and 
corresponding dual one. Physical meaning of the dual invariance
of the black hole entropy is considered in terms of the Euclidean
path integral formulation. 
\end{abstract}
\bigskip
 
\newpage
  
\section{INTRODUCTION}
T-duality implies the equivalence of two apparently very 
different non-linear $\sigma$-models and their respective
corresponding low energy effective theories [1].
This symmetry is described in the context of toroidal
compactifications : for $d$-dimensional compactifications
the T-dual transformation is an element of an infinite 
order discrete symmetry group $O(d,d;Z)$ [2].
It has been also interpreted in terms of the gauging of
an isometry symmetry [3].
For the simplest case of a single compactified dimension 
of radius $R$ on the flat spacetime background, the entire 
physics is left unchanged under the replacement $R 
\rightarrow \alpha^{\prime} /R $ provided one also 
transforms the dilaton field $\phi \rightarrow \phi - 
\log (R/\sqrt{\alpha^{\prime}})$. Thus, geometric properties
and/or topology of the dual\footnote{Hereafter, we use the
word `dual' as only `T-dual'} solution can in general be quite
different from those of the original solution. For example,
it has been shown [4] that the dual solution of the BTZ black
hole [5] is the three-dimensional charged black string solution
[6]. The BTZ black hole does not have a curvature singularity, 
while the black string has a timelike singularity.
Moreover, the former is asymptotically anti-de 
Sitter spacetime, while the latter is asymptotically 
flat\footnote{About the change of asymptotic geometry due to
dualities, see also [7].} 

In this respect it is of great interest to study 
behavior of physical quantities depending 
on geometry of a given spacetime under the 
dual transformation. Horne {\it et al.} [8] 
have shown that on asymptotically flat 
solutions, the duality of the conserved quantities 
defined on asymptotic region is given in such 
a way that mass is unchanged, while the
axion charge and angular momentum are 
interchanged each other.  It has been also 
shown that the Hawking temperature and 
horizon area (viz. entropy) of black hole 
solutions to the low energy effective string theory are
dual invariant [9]. It is a surprising observation
because they are explicitly associated with a 
given spacetime geometry.

As mentioned above, asymptotic behavior of a spacetime 
may be changed by the dual transformation.
Thus, asymptotic flatness is in general not an
appropriate assumption for the study of the duality of 
the physical quantities defined on a given spacetime. 
To avoid this difficulty, we shall introduce a finite
spatial boundary and study the duality of quasilocal
quantities irrespective of asymptotic behavior of
a given spacetime. Quasilocal boundary has been
considered on several studies of gravitational systems
[10 - 18]. Especially, it has been shown that
under the consideration of the quasilocal boundary,
a black hole can be consistently 
interpreted within the framework of the ordinary 
thermodynamics, i.e., the black hole partition function
is well-defined and positivity of the heat capacity
is recovered. [12, 13].

In this paper, we shall consider the duality of the quasilocal
black hole thermodynamics, explicitly the quasilocal black
hole thermodynamic first law, for the three-dimensional low
energy effective string theory. For the purpose, we study
the duality of quasilocal thermodynamic quantities as well as
the black hole entropy on a general axisymmetric three-dimensional
spacetime, say a minisuperspace model [19]. We shall also
consider the physical meaning of the dual invariance of the
black hole entropy in terms of the path integral formulation
[15, 20].

In Sect. II, the quasilocal black hole thermodynamics is briefly
recapitulated. Then, we shall study the duality of the quasilocal
thermodynamic first law in Sect. III. Physical meaning of the dual
invariance of the black hole entropy is also considered in terms
of the Euclidean path integral formulation. Summary and discussions
are given in
Sect. IV.

\section{Quasilocal Black Hole Thermodynamics}
In this section, we briefly recapitulate the quasilocal 
black hole thermodynamics for the three-dimensional 
low energy effective string theory [13, 14]. Quasilocality
means that the three-dimensional manifold $M$ has a 
finite timelike spatial boundary $\Sigma^r$ as well 
as two spacelike boundaries (initial and final ones
denoted by $\Sigma_{t^{\prime}}$ and 
$\Sigma_{t^{\prime \prime}}$, respectively).
The boundary of $\Sigma_t$, which is denoted by $S^r_t$,
is given by an intersection space of $\Sigma_t$ and 
$\Sigma^r$. We assume that $\Sigma_t$ is orthogonal to
$\Sigma^r$. The orthogonality means that on the 
boundary $\Sigma^r$, the timelike unit normal
$u^a$ to $\Sigma_t$ and the spacelike unit
normal $n^a$ to  $\Sigma^r$ satisfy the relation
$u^a n_a|_{ \Sigma^r} =0 $. The induced metric forms
defined on $\Sigma_t$, $\Sigma^r$ and $S^r_t$ are
denoted by $h_{ab}$, $\gamma_{ab}$ and $\sigma_{ab}$,
respectively.

Consider the three-dimensional low energy 
effective string action [6] given by
\begin{eqnarray}
I & = & \frac{1}{2\pi} \int_{M}d^3x \sqrt{-g}
      \Phi \left[R+ \Phi^{-2} (\nabla \Phi )^2
     -\frac{1}{12}H^2 + \frac{4}{l^2}\right]
\nonumber \\ 
    && - \frac{1}{\pi} \int_{\Sigma_{t^{\prime}}
}^{\Sigma_{t^{\prime\prime}}} 
d^2x \sqrt{h} \Phi K 
     - \frac{1}{\pi} \int_{\Sigma^r} d^2x \sqrt{
-\gamma} \Phi \Theta~,
\end{eqnarray}
where $(-1/2)\ln\Phi$ is the dilaton field and $H$ denotes
the three-form field strength of the antisymmetric two-form
field $B$. Comparing with the three-dimensional
general relativity, $-l^{-2}$ can be interpreted as
the negative cosmological constant $\Lambda$ [4]. 
In boundary terms of eq.(1), $K$ and $\Theta $
are traces of extrinsic curvatures of $\Sigma_{t}$ and
$\Sigma^r$ as embedded in the three-dimensional spacetime
$M$, $K_{ab}=-h_{a}^{c}\nabla_{c} u_{b}$ 
and $\Theta_{ab}= -\gamma_{a}^{c} \nabla_{c}  
n_{b}$, respectively. The boundary terms are involved 
such that when one applies a solution of equations
motion into the action and requires the boundary 
condition that field variables be fixed on the 
boundaries, the action has an extremum value.

The canonical form of the action (1) becomes
\begin{eqnarray}
I &=& \int_{M}d^3x \left[P^{ab}{\dot h}_{ab}
 + P_{\Phi}{\dot \Phi} + P_B^{ab} {\dot B}_{ab}
 - N{\cal H} - N^a {\cal H}_a - B_{at} 
\psi^a \right]
\nonumber \\
&& + \int_{\Sigma^{r}}dt d\phi 
 \left[-{\cal E} N + {\cal J}_a N^a - Q ^a 
B_{a t} \right]~,
\end{eqnarray}
where $N$, $N^a$ denote the lapse function and shift vector,
respectively, and ${\cal H}$, ${\cal H}_a$, and $\psi^a$ are 
the Hamiltonian, momentum, and Gauss constraints, respectively.
Conjugate momenta $P$'s are given by
\begin{eqnarray}
P^{ab} &=& \frac{\delta {\cal L}}{\delta {\dot h}_{ab}} =
\frac{\sqrt{h}}{2\pi}\left[
\Phi(h^{ab} K - K^{ab}) - \frac{1}{N}{\dot \Phi}h^{ab} 
+ \frac{1}{N} N^c \partial_c \Phi h^{ab} \right]~,
\nonumber \\
P_{\Phi} &=& \frac{\delta {\cal L}}{\delta {\dot \Phi}} =
\frac{\sqrt{h}}{2\pi}\left[ 2K -  \frac{2}{N}{\dot \Phi}\Phi^{-1} 
+ \frac{2}{N} \Phi^{-1} N^c \partial_c \Phi  \right]~,
\nonumber \\
P^{ab}_B &=& \frac{\delta {\cal L}}{\delta {\dot B}_{ab}} =
\frac{\sqrt{h}}{4\pi N} \Phi h^{ac}h^{bd}\left[
{\dot B}_{cd} + \partial_c B_{dt} + \partial_d B_{tc} \right]~.
\nonumber
\end{eqnarray} 
The integrands of boundary term in eq.(2) are interpreted as the
quasilocal surface energy density ${\cal E} $, momentum density
${\cal J}_a $, and axion charge density $Q^a$ [10, 13, 14]
given by
\begin{eqnarray}
 && {\cal E} = - \frac{\sqrt{\sigma}}{\pi}
\left(n^a \nabla_a \Phi -\Phi k \right),
 \nonumber \\
&& {\cal J}_a =  \frac{2\sqrt{\sigma}}{\sqrt{h}} 
n_c \sigma_{ad} P^{cd}, 
\nonumber \\
&&Q^a =  \frac{2\sqrt{\sigma}}{
\sqrt{h}}P_B^{ab}n_b ,
\end{eqnarray}
where $k$ is the trace of the extrinsic curvature
as embedded in $\Sigma_t$, $k_{ab} =
-\sigma^c_a D_c n_b$, and $D_c$ is the covariant
derivative on $\Sigma_{t}$. These quantities are 
called extensive variables which are composed 
by intensive variables, e.g., the lapse function 
and shift vector. 

According to the Hamilton-Jacobi type analysis, the
informations of a gravitational system with a spatial
boundary are encoded on the boundary and 
variations of the boundary variables determine a
generating functional. Varying the action (2), the
$\Sigma^r$ boundary terms are given by
\begin{eqnarray}
\delta I |_{\Sigma^r} = \int_{\Sigma^r} dt d\phi 
\left[
-{\cal E} \delta N + {\cal J}_a \delta N^a 
- Q^a \delta B_{a t} + N 
\left(
 (s^{ab}/2) 
\delta \sigma_{ab} + {\cal Y}\delta\Phi
\right)
\right],
\end{eqnarray}
where $s^{ab}$, ${\cal Y}$ are interpreted as the
quasilocal surface stress density and dilaton 
pressure density [10, 13, 14], respectively, defined by
\begin{eqnarray}
s^{ab} &=& \frac{\sqrt{\sigma}}{\pi}\Bigl[
\sigma^{ab}n^c \nabla_c \Phi + \Phi [k^{ab}
-\sigma^{ab}(k - n^c a_c)] \Bigr],
\nonumber \\
\cal{Y} &=& \frac{\sqrt{\sigma}}{\pi} \Bigl[
\Phi^{-1} n^c \nabla_c \Phi -(k - n^c a_c) \Bigr]
\end{eqnarray}
where $a^c = u^a \nabla_a u^c = N^{-1} h^{ac}
\nabla_a N $ is the acceleration of the timelike unit 
normal $u^c$. Note that since our considering manifold is
not the four (or higher) dimensional spacetime, but the
three-dimensional spacetime, the term of the electromotive
force is not included in eq.(4) in contrast to the case of
Ref.[14]. These densities (5) appear 
due to the fact that we choose the finite spatial 
boundary, and become zero at asymptotic region
for the asymptotically flat case.

Note that the first three terms eq.(4)
involve variations of intensive variables with 
extensive coefficients. In this paper, it is
appropriate for our purpose to choose the microcanonical
boundary condition in which the thermodynamical 
extensive variables are fixed on the 
boundary [12, 14]. The microcanonical action
can be obtained from the action via the Laplace
transformation as follows
\begin{eqnarray}
I_{micro} &=& I + \int_{\Sigma^{r}}dt d\phi 
 \left[{\cal E} N - {\cal J}_a N^a + Q ^a 
B_{a t} \right]
 \nonumber \\
&=& 
\int_{M}d^3x \left[P^{ab}{\dot h}_{ab}
 + P_{\Phi}{\dot \Phi} + P_B^{ab} {\dot B}_{ab}
 - N{\cal H} - N^a {\cal H}_a - B_{at} \psi^a \right].
\end{eqnarray}
Then, varying the microcanonical action (6), the
$\Sigma^r$ boundary terms are
\begin{eqnarray}
\delta I_{micro} |_{\Sigma^r} = \int_{\Sigma^r} dt 
d\phi N \left[\delta {\cal E} - \omega^a \delta 
{\cal J}_a  + V_a \delta Q^a + (s^{ab}/2) 
\delta \sigma_{ab} + {\cal Y}\delta\Phi \right],
\end{eqnarray}
where $N\omega^a = N^a $ and $NV_a = B_{at}$. 

From the path integral point of view, 
physical states are labeled by the boundary 
variables ${\cal E}$, ${\cal J}^a$ and $Q_a$, and 
parameterized by the boundary metric $\sigma$ 
and the dilaton field $\Phi$ [15]. Thus,
the entropy is statistically defined as the 
logarithm of the density of states 
$\nu({\cal E}, {\cal J}, Q ; \sigma, \Phi)$, which
is a functional of the boundary variables defined
on the timelike spatial boundary $\Sigma^r $.
The density of states is given by tracing the 
microcanonical density matrix $\rho_m (
h_{t^{\prime \prime}}, \Phi_{t^{\prime \prime}},
B_{t^{\prime \prime}}| h_{t^{\prime}}, \Phi_{t^{\prime}},
B_{t^{\prime}} ; {\cal E},
{\cal J}, Q ; \sigma, \Phi)$ from the initial spacelike 
boundary $\Sigma_{t^{\prime}} $ to the final one
$\Sigma_{t^{\prime \prime}}$ as follows
\begin{eqnarray}
\nu({\cal E}, {\cal J}, Q ; \sigma, \Phi) = 
\int {\cal D} [h_t, \Phi_t , B_t] \rho_m (
h_{t^{\prime \prime}}, \Phi_{t^{\prime \prime}},
B_{t^{\prime \prime}}| h_{t^{\prime}}, \Phi_{t^{\prime}},
B_{t^{\prime}} ; {\cal E},
{\cal J}, Q ; \sigma, \Phi),
\end{eqnarray}
where the subscript `$t$' of fields denotes that they 
are not the field variables on the timelike boundary 
$\Sigma^r$, but on the spacelike hypersurface $\Sigma_t$.
The microcanonical density matrix $\rho_m$ 
is given by
\begin{eqnarray}
\rho_m (
h_{t^{\prime \prime}}, \Phi_{t^{\prime \prime}},
B_{t^{\prime \prime}}| h_{t^{\prime}}, \Phi_{t^{\prime}},
B_{t^{\prime}} ; {\cal E},
{\cal J}, Q ; \sigma, \Phi)
= \int {\cal D}[g, \Phi , B] e^{-I^E_{micro}[g, \Phi , B]},
\end{eqnarray}
where $I^E_{micro}$ is the microcanonical
Euclidean action obtained from the microcanonical
action (6) via the Wick rotation $t \rightarrow \tau = i t$.
Then the entropy of a gravitational system 
is given by up to zeroth-order
\begin{eqnarray}
S \approx \ln \nu ({\cal E}, {\cal J}, Q ; \sigma, \Phi)
\approx - I^E_{micro}|_{cl},
\end{eqnarray}
where the subscript $cl$ means that a solution to equations
of motion is plugged into the action. Assuming stationarity
of the solution, the entropy (10) becomes zero. However, if it is
a black hole solution, the entropy does not vanish.

For an observer who lives on asymptotic region, 
all informations are accreted to the event horizon, 
and the horizon can be considered as another boundary,
say inner boundary. However, it must be emphasized that
the inner boundary is not a system boundary on which the
thermodynamic data must be specified [14, 15]. Then the
Euclidean microcanonical action 
has another boundary term as follows 
\begin{eqnarray}
I^{E,~bh}_{micro} 
= I^E_{micro} + \frac{1}{\pi}\int_{\Sigma^{r_H}} 
d\tau d\phi N\sqrt{\sigma}
[\Phi \Theta - n^a \partial_a \Phi],
\end{eqnarray}
where $r_H$ denotes the horizon. As a result, the black
hole entropy becomes
\begin{eqnarray}
S_{BH} \approx - I^{E,~bh}_{micro}|_{cl} 
= - \frac{1}{\pi}\int_{\Sigma^{r_H}} 
d\tau d\phi N\sqrt{\sigma}
[\Phi \Theta - n^a \partial_a \Phi]_{cl},
\end{eqnarray}
where it was assumed that the black hole solution is
stationary, i.e., $I^{E}_{micro}|_{cl} =0$. 
Note that if $\Phi =1$, then the expression for the
black hole entropy (12) is equal to that of Ref.[21],
in which authors have evaluated the black hole
entropy on the action conformally transformed into
the Einstein-Hilbert form.

Finally, for the case that a black hole is 
embedded into a finite cavity, the desired thermodynamic
first law can be obtained via variation of the entropy (12)
[13, 14] as follows
\begin{eqnarray}
\delta S_{BH} = \int_{S^r_t} d\phi ~ \beta  
\left[\delta 
{\cal E} - \omega^a \delta {\cal J}_a  
+ V_a \delta Q^a +  (s^{ab}/2) \delta 
\sigma_{ab} + {\cal Y}\delta\Phi
\right],
\end{eqnarray}
where $\beta=\int N d\tau $ is the inverse 
temperature defined on the finite spatial 
boundary $\Sigma^r$.

\section{Duality of Quasilocal Thermodynamic 
Variables}
We are now ready to study the duality of
the first law of the quasilocal black hole thermodynamics (13).
First of all, consider a general axisymmetric solution
to the equations of motion of the action (1) [19], which is
the form as follows
\begin{eqnarray}
ds^2 &=& -N^2(r)dt^2+f^{-2}(r)dr^2+r^2
\left[d\phi+N^{\phi}(r)dt \right]^2,
\nonumber \\
\Phi &=& \Phi(r),~~B_{\phi t}=B_{\phi t}(r),
\end{eqnarray}
where all other components of the antisymmetric 
field $B_{ab}$ are zero. The horizon $r_H$ 
satisfies the relation $N^2(r_H) =0$. 
In the metric (14), we require the regularity of the
antisymmetric field and vanishing shift vector 
at the horizon as $N^{\phi}(r_H) =B_{\phi t}(r_H)=0$. 
The requirement of vanishing shift vector
at horizon is actually the same with doing a
coordinate transformation $\phi \rightarrow
\phi - \Omega_H t$ on a metric including
non-vanishing shift vector at horizon, 
where $\Omega_H $ is the 
angular velocity of the horizon [9]. 
This sustains the regularity of the
antisymmetric field at horizon under the 
dual transformation.

Since the metric (14) and the fields are independent 
of the coordinate $\phi$, the solution has
a translational symmetry in the direction $\phi$.  
Thus there exists a corresponding 
dual solution by means of the dual transformation [1]
\begin{eqnarray}
g^d_{\phi\phi}& = &g^{-1}_{\phi\phi},~~ g_{\phi\alpha}^d 
 = B_{\phi\alpha}g^{-1}_{\phi\phi},
\nonumber \\
g^d_{\alpha\beta}& = &g_{\alpha\beta}-(g_{\phi\alpha}
g_{\phi\beta}
- B_{\phi\alpha}B_{\phi\beta})
g^{-1}_{\phi\phi},
\nonumber \\
B_{\phi\alpha}^d &=& g_{\phi\alpha}g^{-1}_{\phi\phi},~~
B^d_{\alpha\beta} = B_{\alpha\beta}
- 2g_{\phi [\alpha} B_{\beta ] \phi}g^{-1}_{\phi\phi},
\nonumber \\
\Phi^{d} &=& g_{\phi \phi} \Phi  ,
\end{eqnarray}
where $\alpha$, $\beta$ run over all directions 
except $\phi$. After performing the dual transformation 
on the solution (14), the dual one is obtained by
\begin{eqnarray}
   ds^2_d &=& -N^2(r)dt^2 + f^{-2}(r) dr^2
+ \frac{1}{r^2}\left[d\phi + N^{\phi}_d(r)dt\right]^{2}
\nonumber \\
\Phi^d &=& g_{\phi\phi} \Phi~,~~B_{\phi t}^d = N^{\phi}~,
\end{eqnarray}
where $ N^{\phi}_d(r) = B_{\phi t}(r)$. Note that 
the lapse function and $rr$-component of the metric 
are unchanged under the dual transformation.
Thus the position of the horizon and the original unit normals
$u^{a}$ and $n^{a}$ are unchanged under the
dual transformation. On the other hand, the shift vector
and antisymmetric field are interchanged each other.
It gives us an impression for the duality
of the quasilocal thermodynamic variables, i.e.,
the energy density is unchanged, while the axion charge
density and the momentum density are interchanged
each other. 
Note that here, we set the coordinate $\phi$ to a 
compact one such as $\phi$ is periodically identified 
$\phi \sim \phi + 2\pi$. Then dual solutions represent 
the same conformal field theories [8].

Now, we shall consider the duality of the respective
thermodynamic variables in the following subsections.

\subsection{Black Hole Entropy and Temperature}
It has already been shown in Ref.[9] 
that the black hole entropy and temperature are 
dual invariant for $n$-dimensional black string
solutions. In this subsection, we shall reconfirm
their results with direct calculation of eq.(12) for 
the case of the solution (14).

The Hawking temperature $T_H =\kappa_H / 2\pi$ 
can be obtained from the relation
\begin{eqnarray}
\kappa_H^2 =
 - \left. \frac{1}{2} \nabla^a \chi^b \nabla_a
\chi_b \right|_{r=r_H}~,
\end{eqnarray}
where $\kappa_H$ is the surface gravity and 
$\chi^a =\xi^a + \Omega_H \zeta^a$ 
is the Killing vector normal to the horizon. 
$\xi^a$, $\zeta^a$ are the timelike and spacelike Killing
vectors, respectively. Since we have required that the
shift vector at the horizon $N^{\phi}(r_H)$ vanish, so does
the angular velocity of the horizon $\Omega_H$. The Killing
vector $\chi^a$ is then equal to the timelike 
Killing vector $\xi^a$. Plugging the metric (14) 
and its dual one (16) into (17), it can be shown that the 
surface gravity is dual invariant
\begin{eqnarray}
\kappa_H = \kappa_H^d 
= \left. \frac{f}{2N}(N^2)^{\prime} 
\right|_{r=r_H}~,
\end{eqnarray}
where $\prime$ denotes differentiation 
with respect to the radial coordinate $r$. 
On the other hand, since the lapse function 
$N$ is dual invariant, Tolman temperature 
$T_H/N(r) = T(r)$ [22], which is red-shifted 
temperature from the horizon to the finite 
spatial boundary, as well as the Hawking 
one is also dual invariant, $T(r) = T(r)^d$.

Now, consider the black hole entropy (12).
As mentioned above, the expression for the 
black hole entropy (12) is a generalization
of that of the Einstein gravity, which is 
obtained from the eq.(12) as we set $\Phi = 1$.
It can be also shown that
the black hole entropy (12) satisfies the 
perimeter law, which is the 2+1 dimensional 
version of the area law [5, 19], as follows
; the extrinsic curvature $\Theta$ in the 
metric (14) becomes
\begin{eqnarray}
\Theta |_{r=r_H} =-f \left.
                                      \left(
         \frac{N^{\prime}}{N} + \frac{1}{r}  
                                       \right)
                                                \right|_{r=r_H}
                           = - \left.
                                       \left(
           \frac{2\pi}{\beta_H N} + \frac{f}{r}  
                                       \right)
             			\right|_{r=r_H}.
\end{eqnarray}
And the divergence of the dilaton field is
$n^a \partial_a \Phi = f \Phi^{\prime}$. Then, the black 
hole entropy becomes
\begin{eqnarray}
S_{BH} \approx  -\frac{1}{\pi} \int^{2\pi}_{0} d \phi
\sqrt{\sigma} \left[
-2\pi \Phi - Nf \beta_H \left(
\frac{\Phi}{r} + \Phi^{\prime}
\right)
\right]_{r=r_H}.
\end{eqnarray}
Since the second term in eq.(20) becomes zero
because of $N(r_H)=0$, we finally obtain the
desired black hole entropy satisfying the 
perimeter law for the non-minimally coupled
gravity version [23]
\begin{eqnarray}
S_{BH}= 2 \cdot \left.
\int^{2\pi}_0 d \phi \sqrt{\sigma} \Phi
  \right|_{r=r_H}.
\end{eqnarray}
For the dual solution (16), 
through similar calculation with above one, 
we obtain the dual black hole entropy
\begin{eqnarray}
S_{BH}^d= 2 \cdot \left.
\int^{2\pi}_0 d \phi \sqrt{\sigma^d} \Phi^d
  \right|_{r=r_H}.
\end{eqnarray}
Thus the same form of the perimeter law is 
still satisfied after the dual transformation. 
Furthermore, it can be shown that the black 
hole entropy is dual invariant as
\begin{eqnarray}
S_{BH} =S^d_{BH} =  2 \cdot 2\pi r_H \Phi(r_H),
\end{eqnarray}
where we used the relations
$\Phi^d = g_{\phi \phi}\Phi = r^2 \Phi$,
$\sqrt{\sigma^d} = 1/\sqrt{\sigma} =1/r$.
This is always available to all solutions with types
of the minisuperspace model (14). Note that for the case 
of the three-dimensional general relativity, BTZ 
black hole with $\Phi(r) =1$, the entropy (23)
satisfies the ordinary perimeter law $S_{BH}
= 2 \cdot 2\pi r_H $ [5, 19, 24].

\subsection{Quasilocal Energy Density, Momentum 
Density, and Axion Charge Density}
Horne {\it et al.} [8] have shown that 
for asymptotically flat solutions, the duality of 
the conserved quantities defined on asymptotic
region is given in such a way that
mass is unchanged, while the axion charge and 
angular momentum are interchanged each other.
Let us now examine the duality of 
these corresponding quasilocal densities
for the minisuperspace model (14).

From eqs.(3), (14), and (16), the quasilocal 
surface energy density ${\cal E}$,
momentum density ${\cal J}_{\phi}$, 
and axion charge density $Q^{\phi}$ 
and dual ones are given by
\begin{eqnarray}
{\cal E} &=& {\cal E}^d 
= - \frac{f}{\pi}(r\Phi^{\prime}  + \Phi),
\nonumber \\
{\cal J}_{\phi} &=& - Q_d^{\phi} 
= - \frac{r^3 f}{2\pi N}\Phi (N^{\phi})^{\prime},
\nonumber \\
Q^{\phi} &=&- {\cal J}_{\phi}^d 
= \frac{f}{2\pi r N}\Phi (B_{\phi t})^{\prime}.
\end{eqnarray}
Thus the duality of the quasilocal densities defined on the 
finite spatial boundary is equal to the result of Ref.[6], i.e.,
the quasilocal surface energy density is 
unchanged, while the momentum density and 
axion charge density are interchanged each other.

Note that the quasilocal energy is different from the
quasilocal mass $M(r)$ [16, 17], which is a conserved
charge on evolution of spacelike surfaces, given by
\begin{eqnarray}
M(r) = \int_{S^t_r} dx^{n-2}N {\cal E} 
=  - \int_{S^t_r} dx^{n-2}\frac{\sqrt{\sigma}}{\pi}
N \left(n^a \nabla_a \Phi -\Phi k \right).
\end{eqnarray}
From eq.(25), according to the fact that the lapse function
is unchanged under the dual transformation, it is easily
checked that the quasilocal mass
is also dual invariant.

In the first law of the quasilocal black hole 
thermodynamics (13), 
since the lapse function is unchanged, while the 
shift vector and antisymmetric field are 
interchanged under the dual transformation
in eq.(16), the quantities
$\omega^{\phi} = N^{\phi}N^{-1}$ and $V_{\phi} =
B_{\phi t}N^{-1}$ are also interchanged under the
transformation. As a result, 
the first three terms in the first law (13) 
are invariant under the dual transformation,
\begin{eqnarray}
\delta {\cal E} - \omega^{\phi} 
\delta {\cal J}_{\phi}  
+ V_{\phi} \delta Q^{\phi}
=\delta {\cal E}^d - \omega^{\phi}_d 
\delta {\cal J}_{\phi}^d  
+ V_{\phi}^d \delta Q^{\phi}_d.
\end{eqnarray}

For a noncompact spacetime, e.g., the BTZ black hole,
the quasilocal
quantities in (3) are divergent at asymptotic region.
Thus, for the case, one should introduce a background spacetime
in order to obtain the well-defined quasilocal quantities
[10, 11, 13, 17]. The background spacetime can be chosen
such that if there is not a black hole, the 
quasilocal quantities vanish.
We shall give a comment about that and  in Sect. IV.

\subsection{Quasilocal Surface Stress Density 
and Dilaton Pressure Density}
In this subsection, we study the duality of the 
quasilocal surface stress density $s^{ab}$ 
and dilaton pressure density ${\cal Y}$. 
Existence of these quantities are 
originated from the fact that we have concerned with
the finite spatial boundary $\Sigma^r$. Actually,
in the three-dimensional case, the quasilocal surface 
stress density is just the `pressure' 
on the boundary conjugate to the perimeter $2\pi r$
[10, 13, 14].

For the case of the minisuperspace model (14),
pressure terms in (13) become
\begin{eqnarray}
(s^{ab}/2)\delta \sigma_{ab} + {\cal Y} \delta \Phi
&=& \frac{fN^{\prime}}{2\pi r N}\Phi \delta (r^2)
+ \frac{f}{\pi}\left(1+ \frac{rN^{\prime}}{N} \right) \delta \Phi
\nonumber \\
&=&  \frac{fN^{\prime}}{2\pi^2 N}\Phi \delta (2\pi r)
+ \frac{f}{\pi}\left(1+ \frac{rN^{\prime}}{N} \right) \delta \Phi
\nonumber \\
&=& {\cal P}\delta ({\rm perimeter})
+ {\cal P}_{\Phi} \delta \Phi.
\end{eqnarray}
On the other hand,
from the dual metric (16), their dual forms are given by
\begin{eqnarray}
(s^{ab}_d/2) \delta \sigma_{ab}^d 
+ {\cal Y}^d \delta \Phi^d
&=& \frac{r^2 f}{2\pi}\Phi (2+ \frac{r N^{\prime}}{N}) \delta 
\left(\frac{1}{r^2} \right)
+ \frac{f}{\pi r^2}\left(1+ \frac{rN^{\prime}}{N} \right) \delta 
(\Phi r^2)
\nonumber \\
&=& \frac{r f}{2\pi^2}\Phi (2+ \frac{r N^{\prime}}{N}) 
\delta \left(2\pi \frac{1}{r} \right)
+ \frac{f}{\pi r^2}\left(1+ \frac{rN^{\prime}}{N} \right) \delta 
(\Phi r^2)
\nonumber \\
&=& {\cal P}^d \delta ({\rm perimeter})^d
+ {\cal P}_{\Phi}^d \delta \Phi^d.
\end{eqnarray}
In eqs.(27) and (28), there seems to be no conspicuous
relationship between the (dilaton) pressure
(${\cal P}_{\Phi}$) ${\cal P}$ and the dual one
(${\cal P}_{\Phi}^d$) ${\cal P}^d$. However,
after simple calculation, it can be easily shown that 
the work terms, which are the surface
pressure density times the perimeter and the dilaton 
pressure density times the dilaton field, are not separately
invariant under duality, but are invariant only in the above
combination,
\begin{eqnarray}
(s^{ab}_d/2) \delta \sigma_{ab}^d 
+ {\cal Y}^d \delta \Phi^d
&=&  \frac{fN^{\prime}}{2\pi^2 N}\Phi \delta (2\pi r)
+ \frac{f}{\pi}\left(1
+ \frac{rN^{\prime}}{N} \right) \delta \Phi 
\nonumber \\
&=&(s^{ab}/2)\delta \sigma_{ab} + {\cal Y} \delta \Phi .
\end{eqnarray}
Thus, from the point of view of the duality,
the dilaton pressure density ${\cal P}_{\Phi}$ can be indeed
interpreted as a `pressure' one.

As a result, from eqs.(18), (23), (26) 
and (29), the quasilocal black hole
thermodynamic first law (13) is invariant
under the dual transformation (15).
Specially, the dual invariance of the red-shifted
black hole temperature and the black hole
entropy  tells us that the zeroth and the 
second laws of the black hole thermodynamics
are not spoiled by the dual transformation.
Thus, we can say that though the thermodynamic
variables are closely related to geometry
of a considered spacetime, if two different 
geometric spacetimes are related by the 
dual transformation (15), they are described by
the same black hole thermodynamics.

Before ending this section, it seems to be appropriate 
to comment about the physical meaning of the 
dual invariance of the black hole entropy. 
This is the task of the next subsection.

\subsection{Duality on the Path Integral Formulation}
Following Brown's path integral interpretation 
of the black hole entropy [15], the entropy is
derived from a sum over boundary states,
which are labeled by ${\cal E}$, 
${\cal J}_a$ and $Q^a$, and parameterized
by the boundary metric $\sigma $ and the 
dilaton field $\Phi$. Thus, from this viewpoint,
the dual invariance of the black hole entropy
is the invariance according to 
`coordinate' interchange, ${\cal J}_a  
\leftrightarrow Q^a$, i.e., $({\cal E}, 
{\cal J}_a, Q^a) \rightarrow ({\cal E}^d, 
{\cal J}_a^d, Q^a_d)=({\cal E}, 
Q^a, {\cal J}_a)$, and reparametrization, 
$(\sigma, \Phi) \rightarrow
(\sigma^d, \Phi^d)$ in phase space of the boundary
states. Now, we represent this comment with the 
language of the path integral formulation of [15].

Consider a stationary, non-extreme black 
hole which is embedded within a finite cavity,
and its event horizon is the bifurcate 
Killing horizon satisfying the relation
\begin{eqnarray}
\beta |_B = \beta \omega^a |_B =0 ,
\end{eqnarray}
where `$B$' denotes the bifurcate Killing horizon. 
It must be noted that the density of states 
$\nu[{\cal E}, {\cal J}, 
Q ; \sigma, \Phi]$ in eq.(8) is a functional of the
extrinsic variables and the fields on the outer
boundary not the inner boundary, i.e., the Killing
horizon. In addition,
only the data on the outer boundary are specified.
Thus the density of outer boundary states is given
by the path integral
\begin{eqnarray}
\nu [{\cal E}, {\cal J}, Q ; \sigma, \Phi] =
\int {\cal D} [{\cal E}_B, {\cal J}_B, Q_B, 
\sigma_B, \Phi_B]
\mu [\sigma_B, \Phi_B] \nu[{\cal E}, {\cal J}, Q 
; \sigma, \Phi |
{\cal E}_B, {\cal J}_B, Q_B ; \sigma_B, \Phi_B],
\end{eqnarray}
where $\mu [\sigma_B, \Phi_B] $ is the measure 
for the parameters of the inner boundary states. 
In the steepest descents approximation, eq.(31)
yields the condition that variation of the Euclidean 
microcanonical action, with by the additional term 
$\ln \mu $ with respect to
${\cal E}^B$, ${\cal J}_a^B$ and $Q^a_B$
must vanish 
\begin{eqnarray}
\delta \left. I^{E,~bh}_{micro} \right|_B 
&=& - \left. \int^{2\pi}_0 d\phi \beta  \left(\delta 
{\cal E} - \omega^a \delta {\cal J}_a  
+ V_a \delta Q^a + ( s^{ab}/2) \delta 
\sigma_{ab} + {\cal Y}\delta\Phi \right) \right|_B
\nonumber \\
&& + \left(
\frac{\delta \ln \mu}{\delta \sigma_{ab}}
\delta \sigma_{ab} + \frac{\delta \ln \mu}{\delta
\Phi} \delta \Phi
\right)_B = 0. 
\end{eqnarray}
The first two terms of the integrands in eq.(32) vanish
due to the relation (30). If we again require that the 
antisymmetric field $B_{ab}$ be regular at the 
bifurcate Killing horizon, $B_{ab}|_B =0$, 
the third term of the integrands vanishes. Finally,
the last two terms should be eliminated by variation terms
of the measure factor. Then we obtain the relations
\begin{eqnarray}
 \left. \frac{\delta \ln \mu}{\delta \sigma_{ab}} 
\right|_B
&=& \left. 2\pi \beta (s^{ab}/2) \right|_B 
= \left.\sqrt{\sigma} \Phi \sigma^{ab}
n^c \partial_c \beta  \right|_B 
\approx \left. 2\pi \sqrt{\sigma}
 \Phi \sigma^{ab} \right|_B ,
\nonumber \\
 \left. \frac{\delta \ln \mu}{\delta \Phi} \right|_B
&=& \left. 2\pi \beta {\cal Y} \right|_B 
=\left. 2 \sqrt{\sigma}n^c \partial_c \beta  \right|_B 
\approx \left. 4\pi \sqrt{\sigma} \right|_B ,
\end{eqnarray}
where we used eqs.(5) and (18).  From these 
relations (33), we can read off the measure factor
as
\begin{eqnarray}
\ln \mu[\sigma_B, \Phi_B] 
\approx 2 \left. \int_0^{2\pi} d \phi
 \sqrt{\sigma} \Phi \right|_B.
\end{eqnarray}
The black hole entropy is then 
\begin{eqnarray}
S_{BH} \approx
\ln \nu[{\cal E}, {\cal J}, Q ; \sigma, \Phi]
\approx \ln \mu[\sigma_B, \Phi_B]
\approx 2 \left. \int_0^{2\pi} d \phi
\sqrt{\sigma} \Phi \right|_B.
\end{eqnarray}
In the above equation, we can see that 
the perimeter law (21) is well recovered. 

Now, consider the dual transformation of the
black hole entropy. Since eq.(30)
and the regularity condition of the antisymmetric
field are still satisfied in the dual transformation
(15) and the first three terms of the integrands in 
eq.(32) is dual invariant (see eq.(26)), the relations
for the measure factor (33) are again recovered
in the dual transformation. Thus the dual
measure factor becomes
\begin{eqnarray}
\ln \mu^d[\sigma^d_B, \Phi^d_B]
\approx 2 \left. \int_0^{2\pi} d \phi
\sqrt{\sigma^d}\Phi^d \right|_B.
\end{eqnarray}
Finally, it can be shown that the dual entropy is
still satisfied the perimeter law and equal to
the original entropy for the case of the 
minisuperspace model (14)
\begin{eqnarray}
S_{BH}^d \approx \ln \mu^d[\sigma^d_B, \Phi^d_B]
\approx 2 \left. \int_0^{2\pi} d \phi
\sqrt{\sigma^d}\Phi^d \right|_B \approx S_{BH}.
\end{eqnarray}
As a result, from the path integral viewpoint, the dual
invariance of the black hole entropy means 
that the measure factor of the inner boundary
states for the parameters $\sigma_B, \Phi_B$ 
is invariant under just reparametrization,
$(\sigma_B, \Phi_B) \rightarrow (\sigma^d_B, 
\Phi^d_B) $ . Actually, the reparametrization
invariance is a common event. Thus, according to 
the above interpretation, the dual invariance of 
black hole entropy seems to be quite natural.

\section{Summary and Discussions}
We have considered the duality of the quasilocal 
black hole thermodynamics, specifically the
first law of the quasilocal black hole 
thermodynamics (13). Motivation of this 
work is originated from the fact that the T-dual
transformation (15) may change the asymptotic behavior
of a spacetime. Thus the assumption of the asymptotic
flatness is unavailable in the context of the dual
transformation. To avoid this difficulty, it is
appropriate for the study of the duality of the black hole
thermodynamics to introduce the quasilocal boundary.

It has been shown that the black hole
entropy is obtained in terms of the extrinsic
curvature in a form similar to that in [21], in which
the authors used the conformally
transformed Einstein-Hilbert action. 
This expression 
satisfies a version of the area law that is relevant to
the three-dimensional, non-minimally coupled gravity.
Ii has been also shown that the Hawking temperature
and the black hole entropy are dual invariant, respectively.
In addition, the Tolman temperature,
which is redshifted temperature from the horizon
to the spatial boundary is also dual invariant.

On the other hand, the duality of 
the quasilocal surface energy
density ${\cal E}$, momentum density
${\cal J}_a$, and axion charge
density $Q^a$, which are the quasilocal thermodynamic 
extensive variables in the thermodynamic first law, 
is equal to the result of Ref.[8]; the energy density 
is unchanged, while the momentum density and charge density
are interchanged each other under the dual transformation. 

Since we have considered a quasilocal boundary, the
work terms have appeared in the first law (13),
which are given by the quasilocal pressure densities
times the variations of corresponding conjugate
variables ; in our case, the surface
pressure density times the variation of
perimeter and the dilaton pressure density times
the variation of dilaton
field. It has been shown that the pressure
densities do not have any well-defined dual
behavior, however, the combination of the work terms
is still dual invariant.

The above observations turn out that
the first law of the quasilocal black hole thermodynamics
is dual invariant. In addition, according to the fact that
the Tolman temperature and black hole
entropy are dual invariant, the zeroth and the second laws
are also dual invariant. As a result, though the thermodynamic
variables are closely related to geometry
of a spacetime, if two different 
geometric spacetimes are related by the 
dual transformation, they are described by
the same black hole thermodynamics.
Furthermore, this dual invariance is
irrelevant of the asymptotic behavior
of the spacetimes, e.g., one is asymptotically
flat and the other asymptotically anti-de Sitter.

On the other hand, we have also considered the 
dual invariance of the black
hole entropy in terms of the Euclidean
path integral formulation. Through this analysis,
one can deeply understand a physical 
meaning of the dual invariance as follows;
the dual invariance of
the black hole entropy is just the invariance of the
`coordinate' interchange, ${\cal J}_a  
\leftrightarrow Q^a$, i.e., $({\cal E}, 
{\cal J}_a, Q^a) \rightarrow ({\cal E}^d, 
{\cal J}_a^d, Q^a_d)=({\cal E}, 
Q^a, {\cal J}_a)$ and the reparametrization, 
$(\sigma, \Phi) \rightarrow (\sigma^d, 
\Phi^d)$, in the phase space of the boundary
states. In short, the measure factor 
of the inner boundary states for the 
parameters $\sigma_B, \Phi_B$ 
is invariant under reparametrization.
Thus, in the context of the  path integral
formalism, the dual invariance of the
black hole entropy is quite natural.

For a noncompact geometry, the quasilocal quantities
in eqs.(3) and (5) are not well defined
in the limit $r \rightarrow \infty$, i.e., divergent. 
However, the unexpected divergence can be
naturally eliminated by introducing a reference
background spacetime with an action  $I_0$ 
and being defined a physical action as 
$I_p \equiv I - I_0$ [10, 11]. Then, the physical quasilocal
thermodynamic variables are given as subtracting 
those of the reference background from the 
prescribed ones. In the case, the duality of 
the physical asymptotic thermodynamic variables
as well as quasilocal ones is the same figure 
with that of our consideration in this paper. 
Thus it can be confirmed that the black hole 
thermodynamics in asymptotic region
on the noncompact geometry is still dual 
invariant [18].

Recently, constant curvature black holes, 
$n$-dimensional generalizations of the 
BTZ black hole, and its thermodynamics 
has been carefully studied [25, 26]. Particularly, 
the black hole has an unusual property 
in the black hole thermodynamics
such as the black hole entropy is not given
by the outer horizon instead inner horizon
[25] or depends on the 
size of the quasilocal surface [26]. 
As mentioned above, the duality transformation
drastically changes the geometry of a 
spacetime. Thus, one may inquire as to how the entropy
behaves under such a transformation. It would be
interesting to study the duality of black hole thermodynamics
for $n$-dimensional constant 
curvature black holes.

\section*{Acknowledgments}
We were supported in part by Basic Science 
Research Institute Program,
Ministry of Education, Project No. BSRI-97-2414. 
One of us (W.T. Kim) was supported in part
by the Korea Science and Engineering Foundation 
through the Center for Theoretical Physics in Seoul
National University (1998).

\end{document}